\documentclass[twocolumn,aps,preprintnumbers,amsmath,amssymb]{revtex4}
\usepackage{graphicx}

\begin{document}

\title{Dipole-exchange spin waves in Fibonacci magnetic multilayers}
\author{J. Milton Pereira Jr. and R. N. Costa Filho}
\affiliation{Departamento de F\'{\i}sica, Universidade Federal do
Cear\'a, Caixa Postal 6030\\ Campus do Pici, $60451-970$
Fortaleza, Cear\'a, Brazil}

\begin{abstract}
A microscopic model is employed to calculate the spectrum of spin
waves in quasiperiodic magnetic multilayers in the dipole-exchange
regime. Results are presented for structures in which thin
ferromagnetic films  are separated by non-magnetic spacers following a
Fibonacci sequence and extend  previous magnetostatic calculations.
The results show the splitting of the frequency bands and the mode
mixing caused  by the dipolar interaction between the films as a
function of spacer thickness, as well as the fractal aspect of the
spectrum induced by the non-periodic aspect of the  structure.
\end{abstract}
\maketitle

%\PACS 75.30.Hx  \sep 73.30.Ds \sep 75.30.Pd
%\end{keyword}
%\end{frontmatter}

\section{Introduction}

The development of ultrathin film deposition techniques has allowed
the production of layered materials with properties that are not found
in bulk media. Layered magnetic structures, in particular, have
attracted a great deal of attention due to their possible use in novel
microelectronic devices \cite{Bland}. The transport properties, as
well as the collective excitations, of  magnetic structures in which
magnetic films are interspersed with non-magnetic  layers were shown
to be quite distinct from those of single films. These new  properties
are a consequence of the fact that the interactions between
microscopic elements of the system can be influenced by the structure
of the  multilayer. Thus, by adjusting the composition and the
thickness of the films, one gains the ability to modify the spectrum
of excitations of the structure. One way of achieving that control is
by changing the thickness of the non-magnetic layers, since that can
alter the strength of the dipolar field between the magnetic films.
Several studies have investigated the effect of this long range
interlayer interaction on  the magnetostatic modes of magnetic
structures  \cite{Mika1,Emtage,Mika2,Grunberg}. Recently, a
microscopic theory was developed for the regime in which the dynamic
aspect of the exchange interaction becomes relevant \cite{Milt1}. In
this dipole-exchange  regime, the dipolar interaction acts to couple
the volume spin wave (SW) modes of the magnetic layers  and thus
causes significant modifications of the SW frequency branches, in
comparison with the results for the magnetostatic case.

So far, the microscopic dipole-exchange SW theory has been applied to
layered media in which the thickness of the spacers is assumed to be
constant  throughout the structure. On the other hand, several studies
have focused on the properties of layered structures in which the
width of the elements is varied along its thickness. The interest on
these structures was motivated by reports on the existence of a
quasicrystaline phase in metallic alloys \cite{Blech}, which have been
shown to display long range  orientational order, but are
not invariant under lattice translations. One type of artificial
quasiperiodic structured considered is the Fibonacci multilayer, i.e.
a multilayer with  elements that have widths that vary according to a
binary substitutional sequence  \cite{Merlin,Todd}. The elementary
excitations of these multilayers have been extensively  studied, both
theoretically as well as experimentally (for a review, see
\cite{ELA}).

To date, work on spin excitations in quasiperiodic
magnetic media has been restricted to the magnetostatic regime, in
which the spins of the magnetic layers precess uniformly
\cite{Feng,Dory}.  These calculations, therefore, do not address the
influence of the multilayer structure on  the standing
volume SW modes of each layer, which have amplitudes that vary along
the thickness of the films. The influence of the exchange interaction
is also expected to become  particularly relevant for
structures with ultrathin layers (i.e., with films with thickness of a
few monolayers). Moreover, the dipole-exchange wave vector regime in
thin films has  recently become experimentally accessible, due to the
development of new  experimental techniques that allow the observation
of the SW dispersion across the whole Brillouin  zone \cite{Vollmer}.

In this paper we use the microscopic model presented in Ref.
\cite{Milt1} to the study of magnetic excitations in Fibonacci
multilayers in the dipole-exchange regime. The description  takes into
account the position dependence of the microscopic spins in each
layer. This  model has been applied previously to obtain the spin wave
spectrum of ultrathin films of  ferromagnets \cite{Rai} and
antiferromagnets \cite{Milt2}. The paper is structured as follows: in
section 2 the structure of the multilayer is described and a
microscopic Hamiltonian for the system is introduced. In section 3,
numerical results are presented and discussed for multilayers of
GdCl$_3$ and  EuO. In section 4 the main results are
summarized and a brief conclusion is presented.

\section{Model}

Let us consider a system containing $N_l$ ferromagnetic films
separated by non-magnetic films (spacers), as shown in Fig. 1.
In the following discussion, the ferromagnets and spacers are referred
to as {\it films}, whereas the  spin layers in each ferromagnet are
referred to as {\it atomic layers}. The films have a simple cubic
crystal structure, with lattice constant $a$ and have ideal interfaces
corresponding to  (001) crystal planes. Each of the magnetic films is
a single domain and  contains $N_m$ atomic layers (in the $xy$ plane)
with localized spins, which are assumed to be aligned in-plane and to
interact by isotropic  exchange with its nearest-neighbors and via
dipolar coupling with all other  spins in the structure. All magnetic
films have the same thickness and composition  and every site in the
magnetic lattice has the same number of nearest  neighbors, with the
exception of the uppermost and bottom films, which are assumed to have
vacuum interfaces.

In order to obtain a Fibonacci multilayer, we assume that the
thickness of the spacers can have two different values. Thus, the
non-magnetic spacer films are identified as either $A$ or  $B$,
according to their thickness, which is taken as $d$ for the $B$ films,
whereas  the $A$ spacers have thickness $d + \delta$. The thickness of
the magnetic films, $D$,  is kept constant. The spacers are then
arranged as a binary string, according to the  Fibonacci
sequence. This sequence can be generated by joining the strings
according to the recursive rule ($n\geq 1$):
\begin{equation}
S_{n+1}=\{S_{n}S_{n-1}\},
\end{equation}
where $S_0=B$, $S_1=A$ and $n$ is the generation number.
The strings can also be obtained in an equivalent way by means of the
inflation rule
\begin{equation}
B\rightarrow A, \qquad A\rightarrow AB.
\end{equation}
The total number of elements $A$ and $B$ in each sequence is equal
to a number $F_n$, which is given by the
recurrence formula $F_{n+1}=F_n+F_{n-1}$,
with $F_0=F_1=1$. In the limit $n\rightarrow \infty$ the ratio
$F_n/F_{n-1}$ approaches a characteristic number $\phi =
 (1+\sqrt{5})/2$,
which is the well known golden mean. The multilayer is constructed in
 such a way that
both the upper and bottom films are magnetic, which means that, if the
 total number of
spacers is $F_n$, the number of ferromagnets is $F_n+1$ (see Fig. 1).

The SW dispersion calculation is based on a Hamiltonian formalism in
which the dipolar coupling between
localized spins is included explicitly.
The Hamiltonian for the system is
\begin{eqnarray}
{\mathcal H} &=& -\sum_{i,j}J_{ij}{\bf{S}_{\it i}}\cdot {\bf{S}_{\it
j}} -g{\mu}_B\sum_{i}{H_{0}S_i^z}\cr
&&+(g\mu_B)^2\sum_{\alpha
,\beta }\sum_{l,m}D_{lm}^{\alpha \beta} S_l^{\alpha }S_m^{\beta
},
\end{eqnarray}
where $J_{ij}$ is the exchange between nearest-neighbor sites $i$
and $j$ in the same film
$H_{0}$ is the Zeeman field, taken as parallel
to the $z-$direction. The last term in the Hamiltonian represents the
contribution of the dipolar interaction. The $D^{\alpha \beta}_{lm}$
 are
the long-range dipolar coupling coefficient between any sites $l$ and
 $m$ in the
same or different films. The $\alpha$ and $\beta$ indices
denote components $x$, $y$ or $z$; $g$ is the Land\'e factor and
$\mu_B$ is Bohr's magneton. The expression for the dipolar factors
is

\begin{equation}
D_{lm}^{\alpha \beta}=\frac{[|{\mathbf r}_{lm}|^2\delta_{\alpha
 \beta}-
3r_{lm}^\alpha r_{lm}^\beta]}{|{\mathbf r}_{lm}|^5},
\end{equation}
where the vector ${\mathbf r}_{lm}={\mathbf r}_{l}-{\mathbf
r}_{m}$ connects magnetic sites in the lattice.
In order to calculate the SW frequencies, one can write down the
equations of motion for the operators in the films. These
are then transformed to a representation involving a two-dimensional
 in-plane
wave vector ${\mathbf k}=(k_x,k_z)$ parallel to the film surfaces. The
resulting system of equations can be solved numerically for the
 frequency
$\omega$ of the modes. The Fourier amplitudes of the dipolar terms can
 be
expressed in terms of rapidly converging summations, as shown in
Ref \cite{Rai}. For multilayers, the expressions for the amplitudes
 are
modified by the presence of the spacers. By increasing the thickness
of the spacers, the dipolar coupling between the ferromagnetic films
 can
eventually become negligible, and the resulting
SW spectrum must then reproduce the results for the single film.
\begin{figure}
\resizebox{0.45\textwidth}{!}{\includegraphics{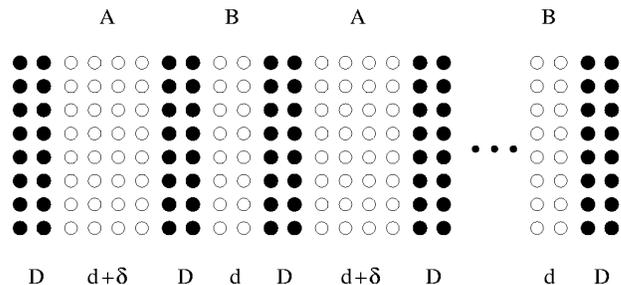}}
\caption{Schematic depiction of a Fibonacci magnetic multilayer. The
the black dots represent ferromagnetic spins, and the white dots are
non-magnetic sites. In this figure, the magnetic films have thickness
$D=2a$, the $B$ spacers have widths $d=2a$ and the $A$ spacers have
thickness $d+\delta$, with  $\delta=2a$.}
\label{fig:1}
\end{figure}
The spin Hamiltonian can be rewritten in terms
of boson creation and annihilation operators by means of the
Holstein-Primakoff transformation \cite{Keffer} which, for low
temperatures (i.e. $T << T_c$, where $T_c$ is the Curie temperature of
the  ferromagnets) can be written as ${S_i}^+ \approx \sqrt{2S}\,
a_i$, ${S_i}^- \approx\sqrt{2S}\, {a_i}^\dagger
$,${S_i}^z = S - {a_i}^\dagger {a_i}$, where ${a_i}^\dagger$ and
$a_i$ are boson creation and annihilation  operators,
respectively.

Next, the equations of motion for the creation and annihilation
 operators can be
obtained, and be transformed to a representation involving a
two-dimensional (2D) in-plane
wave vector $\textbf{k}_n$ = $(k_x,k_y)$, where $n$ is an index
assigned to each {\it magnetic } atomic layer of the system.
The Fourier transforms of the
dipole sums in the Hamiltonian are similar to the terms calculated for
 a
single ferromagnetic film \cite{Rai}, with additional terms
including extra distance factors, which correspond to the dipolar
 interaction
between spins in different films.
In the present case, the different values of thickness of the spacers
 correspond to a
quasiperiodic modulation of the distance factors. This contrasts with
 the previous
theories of Fibonacci multilayers, which were based on transfer matrix
 formalisms.
The present method allows us to introduce the quasiperiodic aspect in
 the structure
in a more straightforward way, by means of a simple modification of
 the dipole sums
expressions of a single film.
Specifically, the expressions for the dipole sums that describe the
 interactions between spins in
different atomic layers contain exponential terms of the type
\begin{equation}
e^{-2|y|\gamma_{ab}},
\end{equation}
where $|y|$ is the distance between the layers, $\gamma_{ab}$ is a
 function of the in-plane
wavevectors and $a$ and $b$ are
dummy indices \cite{Rai}. Thus, for a pair of spins located in atomic
 layers of different
films, an extra distance, corresponding to the thickness of the
 spacers located between said films,
must be added to $|y|$, which would otherwise correspond to the
 distance between the spins if
they were located in the {\it same} film. Thus, when writing the
 expressions for the dipole
sums, one must keep track of the distances between each pair of films.
 If one labels the films
from $1$ to $F_n+1$, the extra separation between two spins in two
 given films with labels
$t$ and $u$ is
%\begin{equation}
%|\Delta(t,u)|=|t-u|(D+d)+L_{t,u}\,\delta.
%\end{equation}
\begin{equation}
|\Delta(t,u)|=|t-u|(D+d)+L(t,u)\,\delta.
\end{equation}
where $L(t,u)$ is a function that gives the number of $A$ blocks found
 between the films
$t$ and $u$, and has the following properties:
$L(t,u)=L(u,t)$, $L(t,t)=0$, $L(t,u)-L(t,u')=L(u,u')$
and (for $u>t$), $L(t,u+1)=L(t,u)+L(t+1,u+1)-L(t+1,u)$.

Having calculated the expressions for the dipole sums, one can write
down the equations of motion for the SW modes in
matrix form. The SW dispersion can then be
calculated numerically as the eigenvalues of a $(2N_l\,N_m)\times
(2N_l\,N_m)$ matrix. %Therefore, the computation time is dependent on
the generation number.

\section{Numerical Results}

Solutions for SW dispersion relations were calculated
numerically for Fibonacci multilayers with parameters corresponding to
the ferromagnet EuO ($T_c \approx 69$ K) and GdCl$_3$($T_c \approx
2.2$ K). The parameters for the dipolar coupling strength (given in
terms of the bulk saturation magnetization) and the exchange field
were, for the GdCl$_3$, $4\pi M = 0.82$ T and $H_{ex}=0.54$ T,
respectively, and $4\pi M = 2.4$ T and $H_{ex}=38$ T for the EuO,
where in both cases $g\mu_BH_{ex}=6SJ$.
\begin{figure}
\resizebox{0.45\textwidth}{!}{\includegraphics{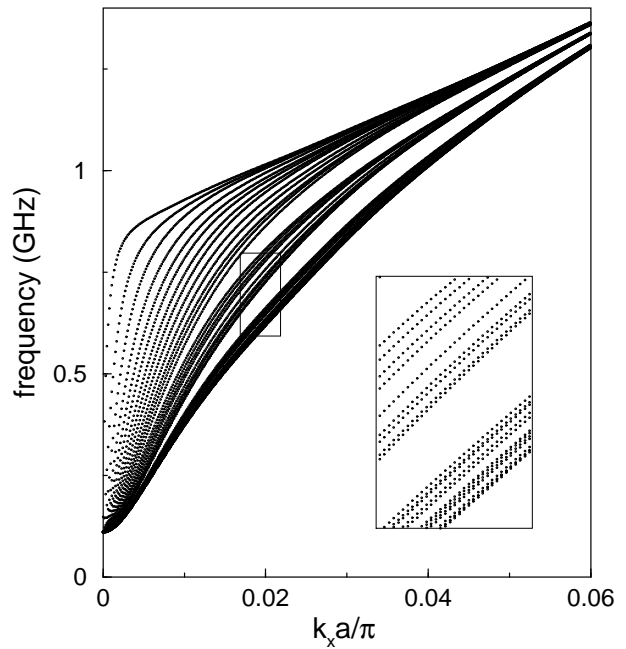}}
\caption{Spin wave dispersion relation for a EuO structure with $35$
 films ($34$ spacers),
for an external field $H_0=0.0$ T. In this case
each ferromagnet has $4$ atomic layers, whereas the spacers have
 thicknesses
$d=20a$ (A) and $5a$ (B), where $a$ is the lattice parameter of the
ferromagnets.}
\label{fig:2}
\end{figure}
\begin{figure}
\resizebox{0.45\textwidth}{!}{\includegraphics{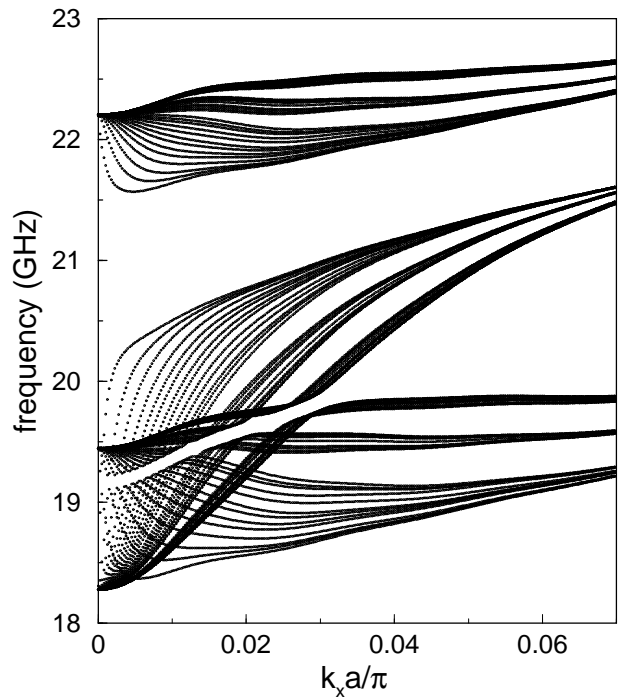}}
\caption{Spin wave dispersion relation for a GdCl$_3$ structure with
 $22$ films ($21$ spacers),
for an external field $H_0=0.36$ T. In this case
each ferromagnet has $5$ atomic layers, whereas the spacers have
 thicknesses
$d=20a$ (A) and $d=5a$ (B), where $a$ is the lattice parameter of the
ferromagnets.}
\label{fig:3}
\end{figure}

Figure 2 shows the lowest SW frequency bands as a function of reduced
 wavevector, for a EuO structure
with $35$ ferromagnetic thin films ($n=8$),
each comprising $4$ atomic layers ($D=4a$),
with the $B$ type spacers having thickness $d=5a$, and the $A$ spacers
 having
$d=20a$ ($\delta=15a$), with zero
external field.
The frequencies were calculated in the
Voigt geometry (i.e. when both the magnetization and the field are
on the plane and $k_y$ = 0) and given in units of GHz (using $\gamma =
 28$
GHz/T). In the absence of the dipolar coupling, the dispersion would
 display four
branches (each consisting of $35$ degenerate modes), with the
 characteristic sinusoidal
dependence on ${\bf k}$. These modes arise due to the quantization of
 the volume modes in
each magnetic film. Since we assumed that the ferromagnets have a
 simple cubic crystal structure,
and also that the strength of the exchange interaction was the same
 for all film layers,
the exchange-dominated regime does not lead to the existence of
 surface SW modes.
As in the periodic case, by introducing the dipolar interaction, the
 volume SW modes
become coupled, which lifts their degeneracy and causes the appearance
 of frequency bands
at small wavevectors. Furthermore, the dipole-dipole coupling causes
 the appearance of
surface SW bands, which result from a widening of the Damon-Eshbach
 modes of a single film.
However, in contrast with the periodic multilayer results, the
 non-periodic aspect of the
present structure causes the appearance of small gaps and the
 splitting of the branches into
sets of subbands. This behavior can be observed in Fig. 2, where the
 inset shows the structure
of minigaps and subbands.
According to previous continuum calculations for thick films, the
 number and distribution
of these subbands tend to form a Cantor set, with a hierarchy of bands
 and
gaps \cite{ELA}. This Cantor set-like structure can also be observed
 in Fig. 2.
A further consequence of the quasiperiodic structure is the appearance
 of isolated modes,
which are found between subbands. One of such isolated modes can be
 identified
in the inset in Fig.2. These isolated branches are usually found in
 quasiperiodic
structures and have been shown previously to correspond to excitations
 that are neither
extended nor localized, but have instead a critical behavior, with
 amplitudes that
display a power-law aspect \cite{ELA}.

The SW dispersion for a GdCl$_3$ structure is shown in Fig. 3. In this
 case, the
multilayer contains $35$ magnetic films and $34$ spacers. The magnetic
 films have
thickness $D=5a$, whereas for the spacers the values $d=20a$ (A
 spacers) and $d=5a$ (B
spacers) were used, for an external field $H_0=0.36$ T.
The figure shows the surface SW band, along with two volume bands. Not
 shown in the figure
are the two higher frequency bands, which display the same pattern of
 gaps and
subbands as the lower ones.
As in the previous figure, the results show the formation of frequency
bands at small wavevectors, with the appearance of small gaps inside
the bands. The present results, however, also show a  strong mixing of
the surface SW band (which corresponds to a broadening of the
Damon-Eshbach modes) and the lowest lying bulk  band. A consequence of
this mixing is the presence of a frequency gap at small wave  vectors,
induced by a mode repulsion effect. This feature has also been
reported for the periodic multilayer case \cite{Milt1}.

\begin{figure}
\resizebox{0.45\textwidth}{!}{\includegraphics{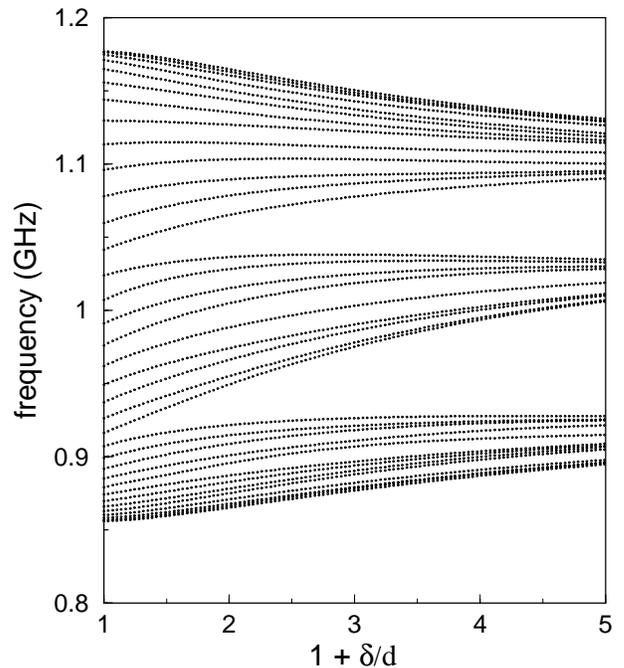}}
\caption{Lowest spin wave frequency branches as a function of the
spacers thickness ratio, for a EuO structure
with $35$ films ($34$ spacers),
for an external field $H_0=0.0$ T and $k_x = 0.03\pi/a$. In this case
each ferromagnet has $5$ atomic layers and the $B$ spacers have
 thicknesses
$d=5a$, where $a$ is the lattice parameter of the
ferromagnets.}
\label{fig:4}
\end{figure}
\begin{figure}
\resizebox{0.45\textwidth}{!}{\includegraphics{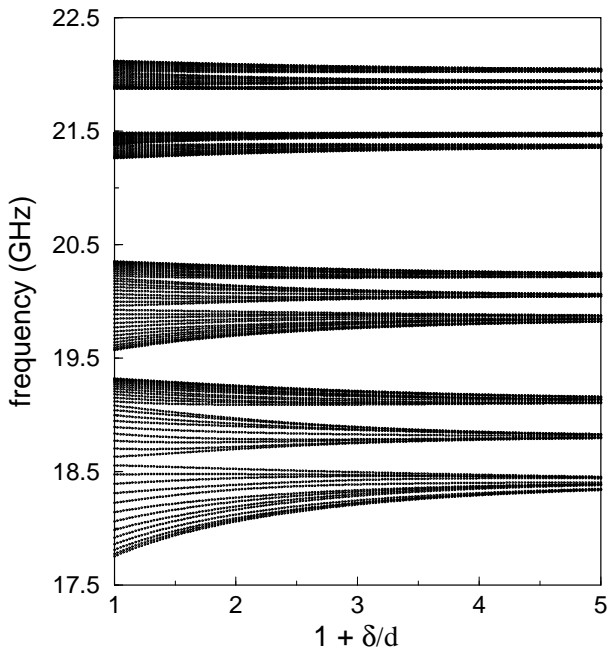}}
\caption{Spin wave frequency branches as a function of the spacers
 thickness ratio,
for a GdCl$_3$ structure
with $35$ films ($34$ spacers),
for an external field $H_0=0.36$ T and $k_x = 0.05\pi/a$. In this case
each ferromagnet has $8$ atomic layers and the $B$ spacers have
 thicknesses
$d=5a$, where $a$ is the lattice parameter of the
ferromagnets.}
\label{fig:5}
\end{figure}

Figure 4 shows a plot of the lowest SW frequencies as a function of
 the thickness parameter
$\delta$ for a EuO multilayer containing $35$ magnetic films, for $k_x
 = 0.03\pi/a$,
$D=5a$ and zero external field. For $\delta = 0$, the results
correspond to the spectrum of a periodic structure with $d=5a$. As the
 thickness of the A
spacers increases, the results show the appearance of gaps and the
 formation
of subbands. A similar behavior is found for a multilayer with $35$
 films of GdCl$_3$, as
shown in Fig. 5, for $k_x = 0.05\pi/a$. In this case the $B$ spacers
 have thickness
$d=5a$, and the ferromagnets have $D=8a$. For the periodic case
 ($\delta = 0$), the results
correspond to $8$ frequency bands, which arise from the $8$ SW modes
 of the films. As the
thickness of the $A$ spacers increases, the figure shows the splitting
 of the
four lowest bands. The same effect is observed in the upper bands.
For larger values of $\delta$, the greater distance between magnetic
 films
separated by the $A$ spacers means that the influence of the dipolar
 interaction across them
becomes negligible, and the results match the SW spectrum of bilayers
 (separated by the
$B$ spacers) and single films (i.e. films located between $A$
 spacers).

\section{Conclusions}
We presented the first results for the dipole-exchange spin wave
 spectrum of
quasiperiodic magnetic multilayers in which thin ferromagnetic films
 are separated
by non-magnetic spacers. The thickness of the spacers was assumed to
 follow
a Fibonacci binary string created by a substitutional rule.
Results were obtained by means of a microscopic
spin Hamiltonian that includes the short-range exchange interaction
 and the long-range
dipole-dipole coupling between localized spins. The dipolar
 interaction, by coupling the
magnetic films, acts to lift the degeneracy of the surface and
 standing volume modes of
the films, thus creating frequency bands, which are distributed as in
 a Cantor set.
The spin wave dispersions also show the presence of isolated frequency
branches that have been known to be associated with critical modes.
These results are consistent
with previous continuum calculations that obtained the long-wavelength
 modes for Fibonacci
multilayers. The presence of the volume modes, due to the position
 dependence of the spin
sites, together with the dipolar interaction, allow the existence of
 mode mixing effects,
which could not be obtained in the continuum calculation.

The authors would like to acknowledge the financial support of the
Brazilian agency CNPq.

\end{document}